# Single electron-spin memory with a semiconductor quantum dot

**Robert J. Young[1], Samuel J. Dewhurst[1,2], R. Mark Stevenson[1], Paola Atkinson[2], Anthony J. Bennett[1], Martin B. Ward[1], Ken Cooper[2], David A. Ritchie[2], Andrew J. Shields[1]**

[1]Toshiba Research Europe Limited, 260 Science Park, Cambridge CB4 0WE, UK
[2]Cavendish Laboratory, University of Cambridge, Cambridge CB3 0HE, UK
E-mail: robert.young@crl.toshiba.co.uk

**Abstract.** We show storage of the circular polarisation of an optical field, transferring it to the spin-state of an individual electron confined in a single semiconductor quantum dot. The state is subsequently readout through the electronically-triggered emission of a single photon. The emitted photon shares the same polarisation as the initial pulse but has a different energy, making the transfer of quantum information between different physical systems possible. With an applied magnetic field of 2 Tesla, spin memory is preserved for at least 1000 times more than the exciton's radiative lifetime.

PACS numbers: 42.50.-p, 42.50.Ct, 73.21.La, 78.67.Hc

Efficient transfer of qubits is a necessity of any useful quantum information application, and requires a powerful link between stored qubits and transmitted quantum data. Photons are an ideal choice for the "flying qubits" required in applications such as quantum cryptography [1] and distributed quantum computing [2]. Quantum information processing however requires qubit operations and storage. Multiple qubit operations are non-trivial with photons as photon-photon interactions are difficult to achieve in practise. Linear optics quantum computing [3] does address this, though photonic memory is still a limiting factor. Conversely atomic and solid state spins [4] are relatively easy to manipulate and store but difficult to transfer over long distances. Storing excitons excited by single photons in semiconductor quantum dots [5] could provide a good interface between flying and stationary qubits. Additionally the semiconductor nature of quantum dots facilitates integration into complex structures which are both compact and robust [6-8], expanding their appeal as an interface between optical and solid-state quantum information. To date a wide range of quantum dot devices have been developed; diode structures to provide electrically triggered single photon emission [7], optical cavities enabling strong coupling between the photonic and excitonic modes [9-11], resonant tunnelling diodes allowing single photon counting [12] and devices capable of emitting polarisation-entangled photon pairs [13-15].

The optical memory we have developed employs a single quantum dot for its active element and is operated as illustrated in Fig. 1. A single quantum dot is embedded in the intrinsic region of a diode. This is biased to make the tunnelling rate of heavy-holes from the quantum dot dominate over the radiative recombination rate of exciton states confined by the dot. A tunnel barrier on the negative side of the diode inhibits electrons from leaving the dot. A weak laser pulse can therefore be used to populate a single electron with a pure spin state into the quantum dot. This is stored until a pair of heavy-holes is returned by the application of an a.c. voltage pulse. Radiative decay of the positively charged exciton follows, leading to the emission of a single photon whose polarisation is determined by the spin-state of the single electron.

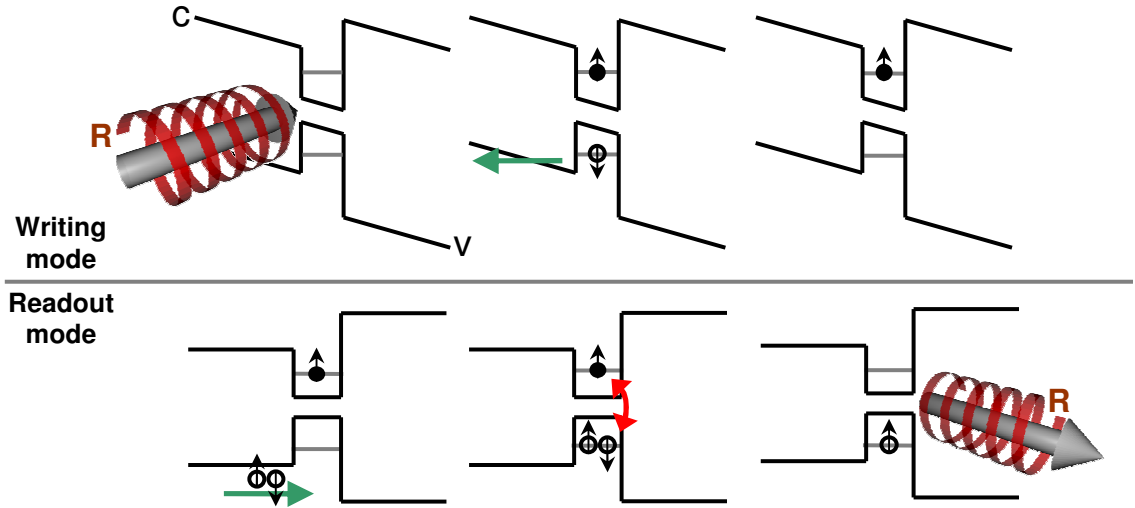

**Figure 1:** Simple schematic diagrams illustrating how the conduction (c) and valence (v) bands around a quantum dot can be manipulated to store single electrons. Writing mode: A single dot is quasi-resonantly excited with a circularly polarised weak pulse of light, the device is biased to remove holes (open circle) from the dot via tunnelling, the polarisation of the pulse is stored on the spin-state of the confined electron (filled circle). Readout mode: The device is biased to return a pair of holes to the quantum dot, radiative recombination results in the emission of a single photon whose polarisation is dictated only by the electron's spin-state. As an example right-hand circularly polarised (R) incident photons are labelled, black arrows illustrate spin orientations.

Fig.2 illustrates the device design we used. Molecular beam epitaxy was used to grow a low density (<1/µm$^2$) layer of InAs quantum dots embedded in GaAs [5]. A 1λ planar optical cavity [16] designed with a mode at ~900nm was formed by 12 repeats of alternating quarter-wavelength thick layers of GaAs/AlAs distributed Bragg reflectors below the dot layer and two repeats of the same mirror structure above the dot layer. The top mirror was p-doped with carbon and the two repeats of the bottom mirror are n-doped with silicon. Electrical contacts to the doped regions allow a field to be applied perpendicularly to the growth plane of the quantum dots. The dot layer was situated on top of a 5nm layer of GaAs above an $Al_{0.98}Ga_{0.02}As/Al_{0.5}Ga_{0.5}As$ superlattice; forming a tunnel barrier, restricting electrons from reaching the n-doped region. An Aluminium shadow mask on top of the sample contains an array of ~2µm diameter circular apertures; these allow single quantum dots to be repeatedly accessed.

The sample was cooled to <10K in a continuous flow Helium-4 cryostat and excited optically with weak ~100ps pulses from a multimode diode laser emitting at 869±5nm. A microscope objective was used to both focus the excitation laser to a ~1µm$^2$ spot on the surface of the device and collect emitted photoluminescence. The photoluminescence was guided into a spectrometer and to a nitrogen-cooled charge-coupled device for wavelength-resolved spectroscopy.

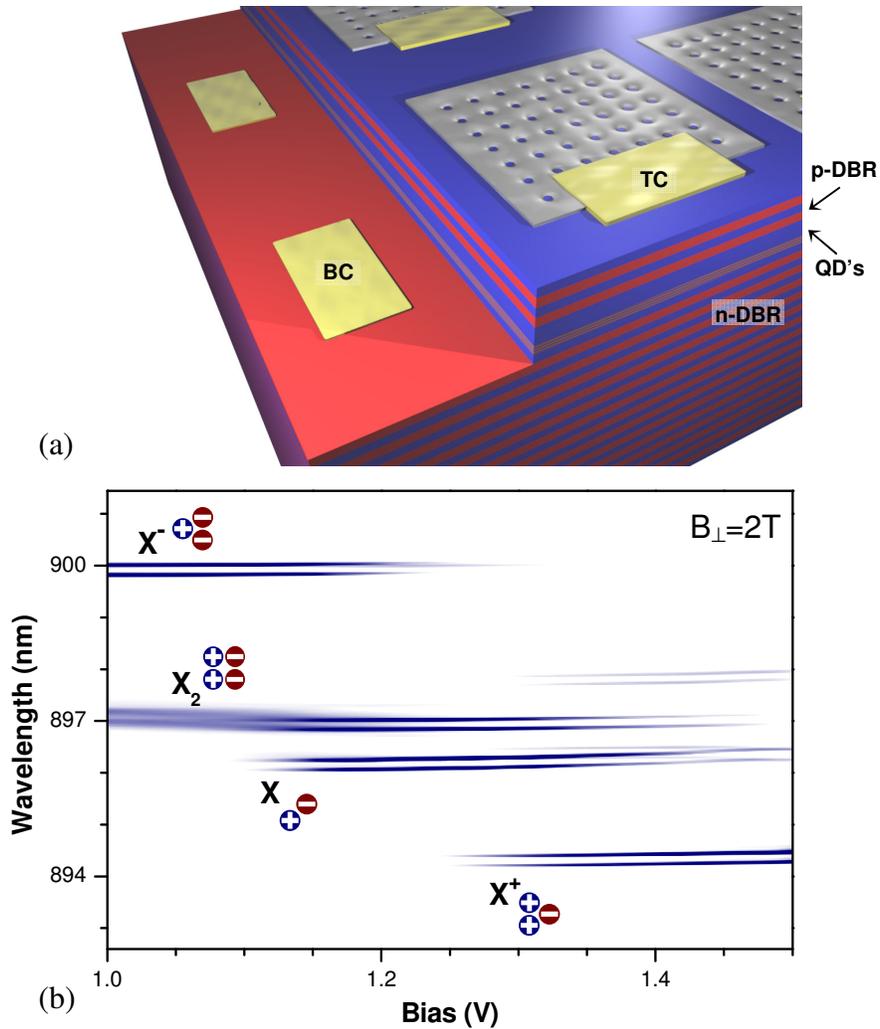

**Figure 2:** (a) An illustration of the sample design. A low density layer of quantum dots (QD's) is embedded in an optical cavity formed by n- and p-doped Distributed Bragg Reflectors (DBR's). Apertures in an aluminium mask allow access to single quantum dots, top (TC) and bottom (BC) electrical contacts are labelled. (b) Photoluminescence spectra taken from a single quantum dot as the bias across the dot is swept. Blue (white) areas indicate high (low) emission intensity. Emission from single positively (negatively) charged $X^+$ ($X^-$) excitons and the neutral exciton X and biexciton $X_2$ are identified. The Zeeman interaction induces an energy separation between states of opposing spin, this is clearly visible for each of the labelled exciton states.

Fig. 2 (b) shows photoluminescence collected from a single quantum dot in the device as a function of the d.c. bias applied to the diode. Emission from the neutral exciton (X, an electron-hole pair) and biexciton ($X_2$, two electron-hole pairs) states is maximised with a d.c. bias of ~1.25V. The dominant charge state can be switched by adjusting the voltage by 0.15V below (above) this bias in which case emission is predominantly found to be from the negatively (positively) charged exciton $X^-$ ($X^+$) states. If a bias is applied which is significantly outside this range then no emission is measured. There is evidence of emission from exciton states containing more carriers, such as the lines visible at ~898nm, though these are generally weak. The application of a 2 T magnetic field perpendicular to the dot layer introduces a Zeeman interaction between states of opposing angular momentum orientation, separating them in energy [17]. This results in the emission of all four of the exciton states, $X^-$, $X$, $X_2$ and $X^+$ splitting into well resolved doublets. The identification of the initial states responsible for the photoluminescence measured from small InAs quantum dots similar to those studied here has been the focus of other studies [18].

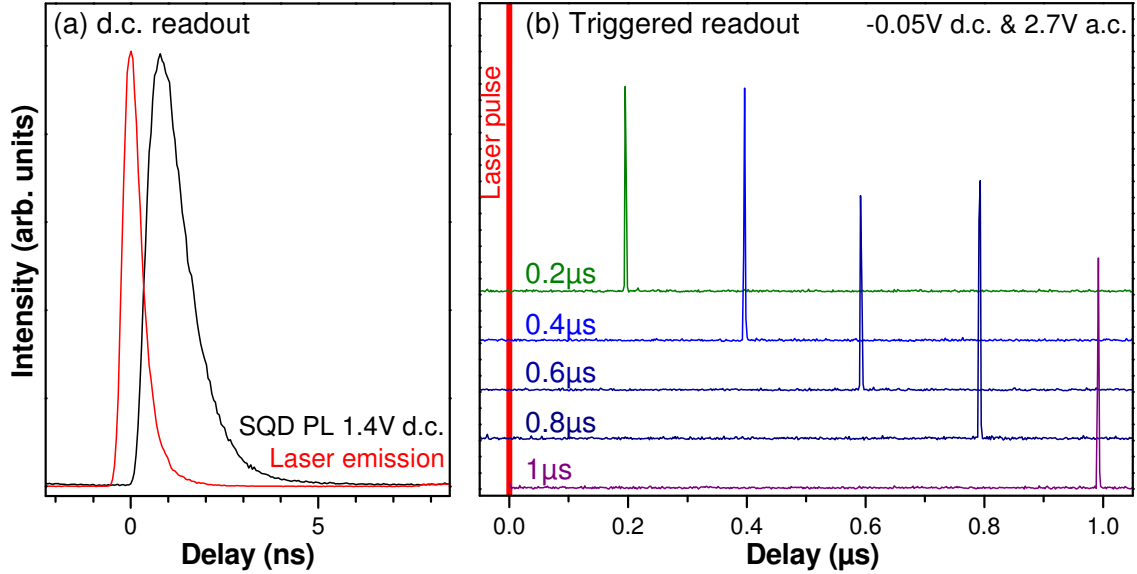

**Figure 3:** (a) Time-resolved emission measured directly from the laser and photoluminescence from the positively charged exciton in a single quantum dot with an applied d.c. bias. (b) Time-resolved emission from the same dot with a -0.05V d.c. bias to remove holes from the quantum dot. 0.2 to 1μs (as labelled) after the laser pulse an a.c. pulse was applied to inject a pair of holes to the dot. Emission from the $X^+$ state is now found to be delayed and in-phase with the a.c. pulse.

To measure time-resolved emission from the $X^+$ state it was selected with a spectrometer and directed into an avalanche photodiode (APD). A single photon counting module measured the time delay between the electronic output of the APD and a trigger pulse from the laser to build a histogram of photon arrival times.

Single electron storage and readout with a quantum dot is demonstrated in Fig. 3. The black line in (a) shows time resolved emission from the $X^+$ state of a single quantum dot. A d.c. bias was applied so that emission from this state was triggered by the laser pulse which was set at a repetition rate of 80MHz. The red trace measures the transient due to the ~100ps laser pulse scattered from the sample's surface. The resolution of this decay curve is limited by the jitter time of the detector used.

Panel (b) in the figure shows emission collected from the same exciton state and dot. The frequency of the laser pulses was lowered to 1MHz and an a.c. voltage pulse was applied on top of a small negative d.c. bias. A pulse-pattern generator was used to both provide a clock signal for the laser used to excite the dot and the a.c. pulses to the sample. The delay between the laser write pulse and the a.c. read pulse was varied from 0.2μs to 1μs. The delay of the observed luminescence correlates with the delay of the a.c. pulse, thus emission is clearly triggered by the readout pulse and not the laser light. The variation in the area of the readout peaks in the traces shown in Fig. 3 (b) is a result of fluctuations in the intensity of the write laser pulse, a result of small lateral movements of the sample in our cryostat.

To confirm that the measured signal can indeed be attributed to electron storage by the quantum dot and is not directly induced electroluminescence by the a.c. pulse, the laser light was removed and no signal was measured (results not shown). Fine control of the applied d.c. and a.c. biases allowed the emission from either the neutral or the positively charged exciton states to be delayed. The latter was chosen for readout as it prevents the formation of long-lived dark states, which would severely limit the speed at which readout can be performed. Limitations in the circuit used to deliver applied a.c. pulse, caused by impedance mismatch between the circuitry and the device, meant the pulse was attenuated somewhat before reaching the device. A larger a.c. pulse was required to put the device into the charge state of a given d.c. bias.

The polarisation of the delayed emission from the $X^+$ state was measured under different pump polarisation conditions. To excite the sample with right- (left-) hand circularly polarised light the linearly polarised laser was passed through a broadband quarter-waveplate placed directly before the microscope objective orientated with its fast axis at +45º (-45º) degrees relative to the laser polarisation. The quarter-waveplate also served to rotate circularly polarised emission from the sample into a linear basis. A half-waveplate and linear polariser were placed directly before the spectrometer, the angle of the latter was fixed to maximise transmission through the spectrometer. Two measurements were taken with the half-waveplate at 0 and +45º; to analyse the emission in independent cross-polarised bases.

With unpolarised excitation each pair of lines shown in Fig. 2 (b) are found, as expected, to be polarised in opposite circular polarisations. When the polarisation of the excitation laser is linear, equal numbers of up- and down-spin electrons are excited in the device and the spin of the electron confined by the quantum dot is random. As a result of this the delayed emission from the two pure-spin components of the $X^+$ state are equal. Fig. 4 (a) demonstrates this, showing equal photoluminescence from both spin states after ~1μs delay following excitation with a horizontally (H) polarised pump. This result is in stark contrast to those shown in Fig. 4 (b) and (c) obtained using a circularly polarised pump, which excites spin-polarised electrons. Here one of the two components of the positively charged exciton is found to be much more intense than the other depending on the orientation of the pump polarisation. The polarisation of the emission correlates with the pump polarisation. These measurements demonstrate that the electron spin is well preserved during the process of relaxing into the quantum dot following non-resonant excitation, removal of the hole from the dot and subsequent re-injection of a pair of holes.

The degree of polarisation memory is defined as $|(I_L-I_R)/(I_L+I_R)|$, where $I_L$ ($I_R$) is the area of the left-hand (right-hand) circularly polarised $X^+$ emission peak. This was found to be 80±10% for both circular pump polarisations and constant within the delay times, 0.2 to 1μs, used in this study. The non-ideal value is thought to derive from a combination of polarisation errors and a degree of spin-scattering as the excited electrons relax into the quantum dot. The longest storage time recorded here is restricted by the repetition rate of the memory, which in turn is limited by the intensity of the emission measured. We expect spin memory to persist for much longer than we can currently measure. Results obtained using a large ensemble of quantum dots have shown that electron spin information can be maintained for times in excess of 1ms [8].

The measured asymmetric lineshape of the exciton emission shown in Fig. 4 is the result of a variable Stark-shift introduced by the applied a.c. pulse. The turn-on time for the pulse is of the same order as the radiative lifetime of the state (~1ns); hence the energy shift induced by the applied electric field varies throughout the readout cycle. The turn-on time of the a.c. pulse is limited here by the equipment used in this study. A difference in intensity of a factor of 2 between the laser triggered output and the electronically triggered readout from the $X^+$ state is observed. This is also predominantly caused by the slow turn-on of the a.c. pulse as it enables some emission from the neutral exciton state, which is not collected in this experiment.

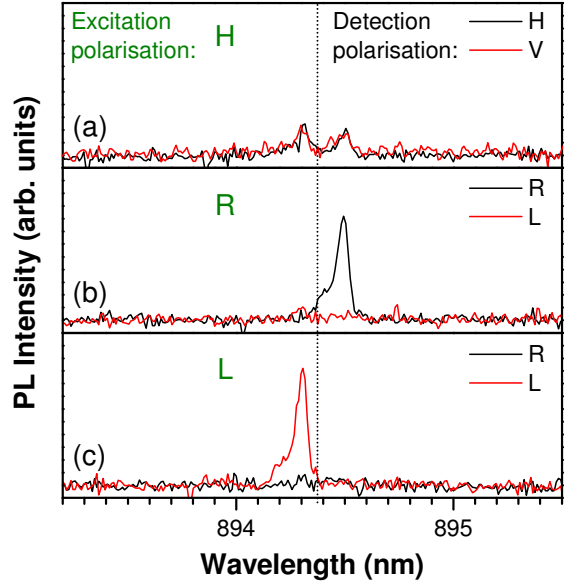

**Figure 4:** Photoluminescence from the positively charged exciton state delayed by ~1μs in a single quantum dot with a 2T perpendicular magnetic field. The three panels show measurements for three differing pump polarisations: (a) Horizontal linear (H), (b) Right-hand circular (R) and (c) Left-hand circular (L). The photoluminescence was measured in cross-polarised bases as labelled, V indicates vertical linear polarisation. The black dotted line is shown as a guide to the eye.

Injecting a pair of holes into the quantum dot as opposed to a single hole prevents the formation of long-lived dark exciton states [17] which could severely limit the readout speed of the device. In this experiment the readout time is limited only by the radiative lifetime of the positively charged exciton. Purcell enhancement [18] of the radiative lifetime of exciton states could provide a means to readout a device operating as demonstrated here on a <100ps time scale. Significant enhancement of the exciton's radiative lifetime has already been demonstrated in systems providing good exciton-cavity mode coupling [19, 20].

The multimode non-resonant excitation used in the write stage here limits the polarisation information transferred from the input pulse to the single electron's spin. Larger quantum dots, providing more confinement than those used in this study, can be quasi-resonantly pumped through a fast relaxing excited state. This should significantly improve the degree of exciton-spin polarisation transferred to the quantum dot. Improvements in the design of the optical cavity should also allow near-deterministic single photon absorption and emission [21]. Single quantum dots could therefore prove to be a useful interface between single photons and electrons finding application in quantum information processing. The difference in energy between the input and output photons in the operation of this device is a very useful feature, it allows the two photons to be spectrally separated and the device to operate with a significant input bandwidth. This could allow the transfer of quantum information between systems operating at different wavelengths.

**Acknowledgements**

This work was partially funded by the EU projects QAP and SANDiE, and by the EPSRC through the IRC for Quantum Information Processing.

**References**

[1] Bennett C H and Brassard G 1984 *Proc. of the IEEE Int. Conf. on Computers, Systems and Signal Processing* 175-9
[2] Cleve R and Buhrman H 1997 *Phys. Rev. A.* **56** 1201-4
[3] Knill E, Laflamme R and Milburn G J 2001 *Nature* **409** 46-52
[4] Monroe C, Meekhof D M, King B E, Itano W M and Wineland D J 1995 *Phys. Rev. Lett.* **75** 4714-7
[5] Bimberg D, Grundmann M and Ledentsov N N 1999 *Quantum Dot Heterostructures*: Wiley, Chichester)
[6] Michler P, Kiraz A, Becher C, Schoenfeld W V, Petroff P M, Zhang L, Hu E and Imamoğlu A 2000 *Science* **290** 2282-5
[7] Yuan Z, Kardynal B E, Stevenson R M, Shields A J, Lobo C J, Cooper K, Beattie N S, Ritchie D A and Pepper M 2001 *Science* **295** 102-5
[8] Kroutvar M, Ducommun Y, Heiss D, Bichler M, Schuh D, Abstreiter G and Finley J J 2004 *Nature* **432** 81-4
[9] Reithmaier J P, Sek G, Löffler A, Hofmann C, Kuhn S, Reitzenstein S, Keldysh L V, Kulakovskii V D, Reinecke T L and Forchel A 2004 *Nature* **432** 197-200
[10] Yoshie T, Scherer A, Hendrickson J, Khitrova G, Gibbs H M, Rupper G, Ell C, Shchekin O B and Deppeet D G 2004 *Nature* **432** 200-3
[11] Peter E, Senellart P, Martrou D, Lemaître A, Hours J, Gérard J M and Bloch J 2005 *Phys. Rev. Lett.* **95** 067401
[12] Blakesley J C, See P, Shields A J, Kardynal B E, Atkinson P, Farrer I and Ritchie D A 2005 *Phys. Rev. Lett.* **94** 067401
[13] Stevenson R M, Young R J, Atkinson P, Cooper K, Ritchie D A and Shields A J 2006 *Nature* **439** 179-82
[14] Young R J, Stevenson R M, Atkinson P, Cooper K, Ritchie D A and Shields A J 2006 *New J. Phys.* **8** 29
[15] Akopian N, Lindner N H, Poem E, Berlatzky Y, Avron J and Gershoni D 2006 *Phys. Rev. Lett.* **96** 130501
[16] Benisty H, Neve H D and Weisbuch C 1998 *IEEE J. Quantum Electron* **34** 1612
[17] Besombes L, Kheng K and Martrou D 2000 *Phys. Rev. Lett.* **85** 425-8
[18] Purcell E M 1946 *Phys. Rev.* **69**
[19] Englund D, Fattal D, Waks E, Solomon G, Zhang B, Nakaoka T, Arakawa Y, Yamamoto Y and Vučković J 2005 *Phys. Rev. Lett.* **95** 013904
[20] Gevaux D G, Bennett A J, Stevenson R M, Shields A J, Atkinson P, Griffiths J, Anderson D, Jones G A C and Ritchie D A 2006 *Appl. Phys. Lett.* **88** 131101
[21] Press D, Götzinger S, Reitzenstein S, Hours J, Löffler A, Kamp M, Forchel A and Yamamoto Y 2007 *Phys. Rev. Lett.* **98** 117402